\documentclass[11pt]{article}

\usepackage{epsfig}
\usepackage{latexsym}
\usepackage{enumerate}
\usepackage{url}
\usepackage{float}
\usepackage[hypertexnames=false,hyperfootnotes=false]{hyperref}
\usepackage{texnansi}
\usepackage{color}
\usepackage{tikz}
\usepackage[margin=10pt,font=small,labelfont=bf]{caption}
\usepackage[subrefformat=parens,labelformat=parens]{subcaption}
\usepackage{afterpage}
\usepackage{enumitem}
\usepackage[boxed]{algorithm}
\usepackage{algpseudocode}
\usepackage[normalem]{ulem}
\usepackage{lmodern}
\usepackage{booktabs}
\usepackage{sectsty}
\usepackage{ifthen}
\usepackage{amsmath}
\usepackage{amsthm}
\usepackage{amssymb}
\usepackage{amsfonts}
\usepackage{thmtools}
\usepackage{thm-restate}
\usepackage{xspace}
\usepackage{titling}
\usepackage{natbib}
\usepackage{xfrac}
\usepackage{multirow}
\usepackage{bigdelim}
\usepackage{bm}
\usepackage[textsize=tiny]{todonotes}
\newcommand{\PR}{\ensuremath{\mathsf{P}}} 
\newcommand{\E}{\ensuremath{\mathsf{E}}} 
\newcommand{\defeq}{\ensuremath{\triangleq}}
\newcommand{\subjectto}{\text{\rm subject to}} 

\newcommand{\Cscr}{\ensuremath{\mathcal C}}

\DeclareMathOperator{\Var}{Var}

\newcommand{\maximize}{\ensuremath{\mathop{\mathrm{maximize}}\limits}}
\declaretheoremstyle[headfont=\sffamily\bfseries,bodyfont=\itshape]{thm-sf}

\renewcommand{\thmcontinues}[1]{\hyperref[#1]{continued}}
\usetikzlibrary{arrows,patterns,plotmarks,pgfplots.groupplots}
\tikzstyle{every picture} += [>=stealth]
\tikzset{axis/.style={semithick, line join=miter}}
\allsectionsfont{\sffamily}
\makeatletter
\def\@seccntformat#1{\csname the#1\endcsname.\quad}
\makeatother

\floatname{algorithm}{\normalfont\sffamily\bfseries Algorithm}
\floatstyle{ruled}
\newcommand{\emailhref}[1]{\href{mailto:#1}{\tt #1}} 
\provideboolean{fastcompile}
\newcommand{\hidefastcompile}[1]{\ifthenelse{\boolean{fastcompile}}{}{#1}}
\usepackage{pgfplots}
\usepackage{pgfplotstable}
\usetikzlibrary{calc}
\usepackage{mathtools}
\definecolor{orange}{rgb}{0.85,0.33,0.13} 
\definecolor{green}{rgb}{0.13,0.85,0.33}
\definecolor{purple}{rgb}{0.33,0.13,0.85}
\definecolor{lime}{rgb}{0.65,0.85,0.13}
\definecolor{blue}{rgb}{0.13,0.65,0.85}
\pgfplotscreateplotcyclelist{tricolor}{%
  orange,every mark/.append style={fill=orange!80!black},mark=*\\%
  green,every mark/.append style={fill=green!80!black},mark=square*\\%
  purple,every mark/.append style={fill=purple!80!black},mark=otimes*\\%
  black,mark=star\\%
  orange,every mark/.append style={fill=orange!80!black},mark=diamond*\\%
  green,densely dashed,every mark/.append style={solid,fill=green!80!black},mark=*\\%
  purple,densely dashed,every mark/.append style={solid,fill=purple!80!black},mark=square*\\%
  black,densely dashed,every mark/.append style={solid,fill=gray},mark=otimes*\\%
  orange,densely dashed,mark=star,every mark/.append style=solid\\%
  green,densely dashed,every mark/.append style={solid,fill=green!80!black},mark=diamond*\\%
}
\pgfplotsset{colormap={tricolormap}{color=(orange) color=(green) color=(purple)},
  colormap={quadcolormap}{color=(orange) color=(lime) color=(blue) color=(purple)}}
\pgfplotstableset{%
  font=\small,
  every head row/.style={before row=\toprule[1pt], after row=\midrule},
  every last row/.style={after row=\bottomrule[1pt]}}
\pgfplotsset{compat=1.15}

\usepackage{setspace}

\title{\textsf{\textbf{The Pricing And Hedging Of Constant Function Market Makers\thanks{The authors would like to acknowledge helpful contributions from Ciamac Moallemi and useful discussions with Agustin Lebron, Tarun Chitra, Guillermo Angeris and Alex Evans.}}}}
\author{
  Richard Dewey \\
  Proven  \\
  email: \emailhref{rdewey@proven.tools}
  \and
  Craig Newbold \\
  Proven \\
  email: \emailhref{cnewbold@proven.tools}
}
\date{Current Revision: June 20, 2023}

\begin{document}
\maketitle
\onehalfspacing

\begin{abstract}
We investigate the most common type of blockchain-based decentralized exchange, which are known as constant function market makers (CFMMs). We examine the the market microstructure around CFMMs and present a model for valuing the liquidity provider (LP) mechanism and estimating the value of the associated derivatives. We develop a model with two types of traders that have different information and contribute methods for simulating the behavior of each trader and accounting for trade PnL. We also develop ideas around the equilibrium distribution of fair price conditional on the arrival of traders. Finally, we show how these findings might be used to think about parameters for alternative CFMMs.  
\end{abstract}

\onehalfspacing

\section{Introduction}
Limit order books have been the dominant trading mechanism in modern financial markets including most global equity and futures markets. Limit orders specify an \textit{amount} willing to be bought or sold at a specified \textit{price}. Alternative market designs such as posted-price where dealers maintain two-sided markets often prevail in bond and foreign exchange markets. 

The explosion of interest in cryptocurrencies has catalyzed interest in decentralized exchanges and automated market makers. Since first introduced in 2018, blockchain-based decentralized exchanges have experienced rapid growth and user adoption. However full limit order books are too computationally costly for smart contracts. Instead, these exchanges employ automated market makers (AMMs), which are set by algorithms and essentially operate as contract-to-peer transactions. 

While many different AMM designs have been proposed, most decentralized exchanges (DeXs) use some flavor of a constant function market maker CFMM, whereby any trade changes the pool such that the product of the pool remains the same. Different variations on this theme are found, but the spirit is similar. CFMMs differ from order books in a significant way. Rather than specifying a price and quantity, market makers make a commitment to participate in a curve. 

The promise of decentralized markets and CFMMs is that they allow any participant to be a market maker and earn the associated fees regardless of their size, which is practically impossible on modern centralized exchanges. The decentralized exchanges also offer a venue for trading crypto assets that might otherwise not meet certain requirements for centralized exchanges. CFMMs also offer an automated method for rebalancing a portfolio of two assets. Finally, CFMMs promise a transparency an alternative incentive structures from centralized exchanges. 

Market making is often seen as a profitable opportunity in traditional markets, but one that requires significant costs and expertise. CFMMs offer the ability to be a liquidity provider (LP), essentially a market making function, according to a constant function that is clear to all participants. LPs are rewarded with fees that are meant to compensate them for their risk which include adverse selection from informed traders. In this paper we present a model for valuing the opportunity to provide liquidity on CFMMs. We show the LP opportunity is akin to selling an option and extend the model to value the relevant derivatives of the option. 

\subsection {Motivations}
Uniswap is the most popular DeX with thousands of crypto-based tokens and hundreds of millions of dollars in notional value traded daily. Uniswap and other CFMMS  such as Balancer, SushiSwap and Curve are the first successful implementations of decentralized exchanges as measured by trade volume, users, longevity and popular interest and turned academic speculation a reality. Subsequently dozens of similar protocols and projects were launched and there is much excitement that they might represent a next step in market design. 

The amount of value being exchanged, combined with the promise of more transparent and fair execution, has attracted wide attention. Moreover, decentralized exchanges and automated market makers are at the core of decentralized finance, which aims to have a broad and significant impact on traditional finance and more modestly positions itself as central to the Web3 and digitally-native economy. 

\subsection {Contributions}
The literature on AMMs is rich and growing by the day. Our contributions are threefold. First, AMMs are often billed as investment products. The idea being that to act as a liquidity provider or market maker, earns a stream of income. Another way to think about these AMMs is like a short option position. Our model is the first to value these pools empirically and provide the relevant greeks. Following convention, our model is calibrated to real world data. 

With our model we then examine the return and risk characteristics of providing liquidity on a delta hedged basis. This has implications for small retail traders and larger market makers who might want to use decentralized exchanges as alternative trading venues. Given the valuation model and greeks, we develop a process for determining fair compensation for providing liquidity. We should note that while fees are the primary source of LPs compensation, yield farming is often an important secondary income component although it's outside the scope of this paper.  We also develop novel accounting methods and techniques around the equilibrium distribution of the fair price for a pool and simulation techniques. Finally, our model also includes fee and a treatment of fees, which makes the math more tractable and will likely be a tool for future researchers.

\subsection {Literature Review}
The literature on AMMs is often traced to \citet{hanson2007logarithmic} who proposed a logarithmic market scoring rule. The scoring rule provides a mapping from prices to pool sizes and has been implemented in prediction markets and online ad auctions. However these mechanism are often complex and although researchers such as \citet{othman2011automated} proposed modifications to make the market design more attractive, these markets still suffer from low participation rates and poor liquidity. 

The 2018 introduction of Uniswap on the Ethereum blockchain sparked a renewed interest. The constraining nature of smart contracts forced innovation and while surprisingly simple, CFMMS have proven to be a robust market mechanism design. Indeed \citet{angeris2019analysis} show how Uniswap tracks the reference price and remains stable under a wide range of market conditions. \citet{angeris2020improved} show how CFMMs can be viewed as alternative price oracles and the authors have written several other papers on different pricing curves and setting the optimal level of fees. \citet{young2020equivalence} shows how AMMs compare to LOBs and the benefits and drawbacks of each mechanism in relation to the other. Finally, \citet{lipton2021automated} explore central bank digital currencies and how it's possible to generate on-chain prices for assets that are consistent with traditional off-chain markets. 

Despite the popular success and interest in decentralized exchanges, they are not without issue. \citet{barbon2021quality} compare decentralized exchanges to centralized exchanges and show the under most conditions decentralized exchanges offer more expensive trade execution. Likewise \citet{capponi2021adoption} shows that being a liquidity provider on most decentralized exchanges has a negative expected value and that participants are likely trading on these venues for alternative reasons. Others such as \citet{aoyagi2021liquidity} show how information asymmetry can lead to issue of toxic flow on decentralized exchanges. 

The community around decentralized continues to evolve and propose improvements. Decentralized exchanges on Solana offer lower costs and structures similar to LOBs. Likewise the latest version of Uniswap (v3) offers many advancement as noted by \citet{White2022}. Ideas such as a time weighted mechanism for trading on CFMMs as suggested by \citet{Robinson2021} also might appear potentially fruitful lines of research. 

We also draw attention to \citet{Milionis2022LVR} which gives a theoretical value of the compensation for being a liquidity provider on a Uniswap pool. The authors examine and develop a model that quantifies adverse selection and provides a framework for valuing the expected returns to being a hedged liquidity provider.  Our work empirically validates the theoretical value that their model produces. We developed a novel methods for modeling the equilibrium distribution of the pool and point to work in \citet{Milionis2023Fees} that offers a complete proof of our method.    

The growing literature on these novel market making design mechanisms is fascinating and impressive. There exists many other papers and research focused on technical topics that is outside the scope of our research, but we include a list of resources we found helpful at the end. We view this blockchain-based decentralized exchanges as a fertile area for further research and innovation. 


\section{Model}

\subsection{Idealized Model}
We propose a model for the price evolution of crypto assets. In particular our model considers trading a risky coin against a numeraire over a finite time horizon to time $T$.  Our model is in continuous time where $t \in [0, T]$ and we use a standard geometric brownian motion (GBM) and a Merton Jump Diffusion (MJD) process, which incorporate jumps. Working in continuous time, which give our model the ability to capture agent arrivals that happen randomly. This price will serve as our reference price on a centralized exchange. \\

\begin{enumerate}
\item Geometric Brownian Motion: 

\[
\frac{{d}{P}_{t}}{{P}_{t}} =  {\mu_{D}}{d}{t} + \sigma_{D} {d}{B}_{t}
\]

\item Merton Jump Diffusion:

\[
  \frac{dp_t}{p_t} =
  \left( \mu_D - \lambda_J k \right)\, dt
  + \sigma_D \, dB_t
  + (y_t - 1) \, dN_t,
\]
where $B_t$ is a standard Brownian motion, $N_t$ is the Poisson process with intensity
$\lambda_J$, and $\log y_t \sim N(\mu_J,\sigma_J^2)$. Here,
\[
k \defeq \E[y_t - 1] = \exp\left(\mu_J + \tfrac{1}{2} \sigma_J^2\right) - 1.
\]
\end{enumerate}

Where $\lambda$ is the average number of jumps per unit of time. We model the log returns of the pool as follows:

\[
  \log \frac{p_{t+\Delta t}}{p_t}
  = \left( \mu_D - \tfrac{1}{2} \sigma_D^2\right) \Delta t
  + \sigma_D \sqrt{\Delta t} Z_0
  - \lambda k \Delta t
  + \sum_{i=1}^N \left( \mu_J + \sigma_J Z_i \right),
\]
where $N\sim \text{Poisson}(\lambda_J \Delta t)$, $Z_i \sim N(0,1)$, for $i \geq 0$.
Conditional on $N$, log returns are normally distributed, so that
\[
  \log \frac{p_{t+\Delta t}}{p_t}
  = \left( \mu_D - \tfrac{1}{2} \sigma_D^2- \lambda k\right) \Delta t
  + N \mu_J
  + \sqrt{\sigma_D^2 \Delta t + N \sigma_J^2} Z,
\]
where $Z \sim N(0,1)$.
It then follows that:
\[
  p_{t+\Delta t} / p_t
  =
  \exp\left(
    \left( \mu_D - \tfrac{1}{2} \sigma_D^2- \lambda k\right) \Delta t
    + N \mu_J
    + \sqrt{\sigma_D^2 \Delta t + N \sigma_J^2} Z
  \right).
\]

The models above are all standard in the finance literature and aside from a few well-known limitations we discuss below give us a flexible and robust structure for the pricing evolution of crypto assets. 

\subsection{Liquidity Pool}
We use a standard model and notation to described the CFMM liquidity pool. The liquidity pool contains the reserves of two tokens, the risky asset and the numeraire. We define the two pieces of the liquidity pool as follows: \\

\noindent{$x_t$ = quantity of reserves in risky asset}

\noindent{$y_t$ = quantity of reserves in numeraire} \\

Without loss of generality the risk-less rate is set to zero. Also without loss of generality our model describes a risky asset and a numeraire rather than two risky assets. This process is right continuous with left limits. The value of the pool at any given time is shown to be: \\

$V_t$ = $x_t$ $p_t$ + $y_t$ \\

We are interested in computing the expectation of $\E[V_T]$. This expected value will serve as a proxy for price. Taking the expectation to be a proxy of price in the Black-Scholes world works because an investor is able to dynamically hedge; the market is said to be complete. In practice this is rarely the case because most traditional assets as well as crypto assets experience jumps and continuous dynamic hedging is not achievable. 

A process with jumps renders the market incomplete and violates a key assumption of the Black-Scholes model. However even though the market is incomplete many models use risk neutral pricing  to value securities, including short-term options which are very sensitive to jumps. Although not perfect, this framework has shown to be effective for decades in the financial economics literature. 

By convention, instead of computing the expected value, we compute the impermanent loss. Computing the expected impermanent loss allows us to easily derive the expected value. Impermanent loss is the difference in value between just holding the time zero reserves and the value of the pool. This can be seen as follows: \\

Impermanent loss: $L_T$  = $x_0$ $p_t$ + $y_0$ - $V_T$  \\

Expected Impermanent loss: $\E[L_T]$  = $x_0$ $p_0$ + $y_0$ - $\E[V_T]$ \\

We assume risk-neutral pricing so $p_0$ = $\E[p_t]$. 

This also assumes that $\mu_D = 0$ and $\mu_J = 0$ as the return on the underlying also must be zero.

\subsection{Agent and Pool Dynamics}

We use two different approaches to model the arrival of agents and the evolution of pool dynamics. In each model, there are two types of traders, arbitrageurs and noise traders. The arbitrageurs know the true fair value of the asset while the noise traders do not. If price discovery happens on a centralized exchange, (as most evidence suggests) then the arbitrageur can reference this price prior to trading. In the first model the noise trader arrives and trades with a random fair value. The second model has the noise traders flipping a coin to determine direction and then choosing a random size according to an exponential distribution. 

In each model, arbitrageurs act instantaneously after each block arrival, and trade optimally around the reference price. Suppose at time $t$ they observe the pool to have reserves ($x_{t-}$, $y_{t-}$). The arbitrageur seeks to trade with the pool to maximize expected profit given the fair reference price. Specifically, the arbitrageur solves the optimization problem:
\begin{equation}\label{eq:arb-action}
  \begin{array}{ll}
    \maximize_{u_x, u_y} & p_{t} u_x + u_y \\
    \subjectto & (u_x,u_y) \in \Cscr(x_{t-},y_{t-}).
  \end{array}
\end{equation}
Here the decision variables $(u_x,u_y)$ determine the quantities of risk asset and numeraire
traded, respectively, against the pool. The set $\Cscr(x_{t-},y_{t-})$ is the set of all trades
permitted by the pool, based on the prior state $(x_{t-},y_{t-})$.
The pool then adjusts according to
\begin{equation}\label{eq:pool-update}
x_t = x_{t-} - u_x, \qquad y_t = y_{t-} - u_y.
\end{equation}

$\Cscr$ is the set of allowed reserves and thus is the set of permissible trades that the pool can transition to after an agent trades. We will define $\Cscr$ below when we define the transition function for Uniswap. 

There are two approaches to modeling noise traders, but in both cases noise traders arrive at times 0 < $t_1$ < $t_2$ < ... and are modeled as a Poisson process with arrival rate ${\lambda_\delta}$. In our first model of noise traders (NT1), an agent arrives at the time $t$ and has an idiosyncratic value of:

\[
	\tilde p_t = p_t (1 + \delta_t).
\]

\noindent $\delta_t$ is a random variable drawn from an i.i.d. lognormal distribution with mean zero and variance $\sigma_{\delta}^2$. $\delta_t$ is an idiosyncratic value offset. 

The trader values the pool by examining the reference price and then adjusting it based on their idiosyncratic preferences and the value of the portfolio. If $\delta_t$ is very small they have very little idiosyncratic preference, they will simply trade if the pool prices deviates from the reference market price. Another way to think about delta is modeling the toxicity of the flow. For example, the smaller the value of $\sigma_{\delta}$, the less the pool is worth, because you have more adverse selection. It can be noted that arbitrageurs are a special case of this model, with $\sigma_{\delta} = 0$.)

They are assumed to be risk-neutral and trade to maximize profit given their idiosyncratic value. This gives an optimization problem similar to that of arbitrageurs, namely:
\begin{equation}\label{eq:agent-action}
  \begin{array}{ll}
    \maximize_{u_x, u_y} & \tilde p_{t} u_x + u_y \\
    \subjectto & (u_x,u_y) \in \Cscr(x_{t-},y_{t-}).
  \end{array}
\end{equation}

In our second model (NT2) of noise traders, the agents flip a coin to determine the direction of the trade. They then determine a random size to trade according to an exponential distribution. Our NT2 model takes the direction and size from the aforementioned distributions and generates the hypothetical trade the agent would make. We feel this model might better reflect reality in that noise traders are coming into trade for without a good sense of the current fair value. There is anecdotal evidence for non-economic trading activity in crypto and in particular on decentralized exchanges. This second model also gives a better fit on the data. 

\[
\tilde p_t = p_{t} \exp(-\sigma_p^2 \Delta t/2 + \sigma_p \sqrt{\Delta t} Z),\quad Z\sim N(0,1),
\]
where $\sigma_p$ is the volatility of the pool valuation. We can write this as
\[
\log \tilde p_t = \log p_{t} -\sigma_p^2 \Delta t/2 + \sigma_p \sqrt{\Delta t} Z,
\]
so that
\[
  \sigma_p = \frac{1}{\sqrt{\Delta t}}\mathrm{stdev}\left( \log \frac{p_{t+1}}{p_t} \right),
\]
this can be used for estimating volatility. Time units (for $\Delta t$) should be daily. \\

This is the realization of the right continuous, left limits process described above. For the sake of clarity, the process works in the following way. The agent arrives at instant $t$. We observe the value of the pool right before $t$ and take the value at time $t$ at the limit and add the value of the trade that the agents elects to execute.  

\subsection{Uniswap Transition Function}
Uniswap is a constant product market maker. More concretely that means that the product of the reserves after any trade should match the product of the reserves prior to the trade. In the transition functions below we set $\theta=1/2$ and $\gamma=(1-0.003)=0.997$ for the Uniswap specific case.  The set of permissible trades that with the constant product condition that defines $\Cscr$ is as follows:

\[
u_x \leq 0, (y-u_y)(x-\gamma u_x) = xy 
\]
\[
u_y \leq 0, (y-\gamma u_y)(x - u_x) = yx
\]

\noindent As shown in [Cite G3M fees App A], the solution to \eqref{eq:agent-action} when $\tilde p_t<\gamma\frac{y}{x}$

\[
u_y = y_{t-}-\left(\frac{\tilde p_t x_{t-} y_{t-}}{\gamma}\right)^{1/2}
\]

\[
u_x = \frac{x_{t-}}{\gamma} - \left(\frac{y_{t-}}{\tilde p_t}\right)^{1/2} \left(\frac{x_{t-}}{\gamma}\right)^{1/2}.
\]

Similarly, when $\tilde p_t>\gamma^{-1}\frac{y}{x}$
\[
u_y = \frac{y_{t-}}{\gamma} - \left(\tilde p_t x_{t-}\right)^{1/2} \left(\frac{y_{t-}}{\gamma}\right)^{1/2}.
\]

\[
u_x =  x_{t-}-\left(\frac{y_{t-1}x_{t-1}}{\gamma\tilde p_t}\right)^{1/2}
\]

where $1-\gamma$ is the fee charged on amount of the assets added to the pool reserves \footnote {The literature on CFMMs sometimes describes this parameterization in terms of L and p as in \citet{Milionis2022LVR}} \\

In the case that $\gamma\frac{y}{x} \leq \tilde p_t \leq \gamma^{-1}\frac{y}{x}$ the agent will decide not trade. Ethereum blocks are minted every 15 seconds, so with this cadence we might observe more than 1 agent per block or no agents in a block. The overall time of the full simulation is four hours, so the expected number of total arrivals would be on the order of 400-500 trades per hour. We run 1000 simulation trials of a one hour trading runs. We believe this is enough arrivals and simulation trials for the results to be robust. 

\subsection{Discussion and Limitations}
We value the investment of a single liquidity provider in the pool. The payoff will be impacted by other providers joining or leaving the pool, particularly if that happens in a way that is correlated with future prices.

The first limitation worth discussing is that time is not continuous, but discrete as these CFMMs are only updated when a new block is minted. A related issue is that there is a meme pool that trades enter before being executed. Our model assumes that an agent arriving to trade achieves instant execution. In reality a trade sits in the meme pool, which causes delays in execution time.

A related issues is that there are non-trivial transaction fees. Specifically on-chain trades must purchase gas which has a highly variable price. Gas price increases are driven by demand and can are correlated with high volatility. Said differently, when arbitrage opportunities look the most compelling, during highly volatile periods, the ability to realize the opportunity might be compromised by the high price of gas. Gas is sold at auctions, which have their own dynamics and complexities. For the purposes of this analysis we ignore gas and other transaction fees, however we acknowledge their potential importance. 

Another element is liquidity provider dynamics. The key insight here is that the profit and loss of a liquidity provider is a function of whether other people decide to leave or join the pool in the future. Our model does not account for this. Instead we look at the pool and assume the amount of liquidity is constant. 

We, and many practitioners, believe that there are likely to be interesting negative gamma dynamics at work in these CFMMs. For example, if a pool suffers too much impermanent loss then LPs will exit the pool, which in turn generates trading that will increases losses. This is in addition to the negative gamma effects observed in a normal function market, which we discussed in the numerical results section.


\section{Empirical Data and Calibration}

In this section we'll give an overview of our data and calibration techniques. For reference trade prices our data comes from Binance, the highest volume crypto exchange. We observe the reference trades every 15 seconds on the pair ETH/USDC, which is one of the most popular crypto pairs. 

For Uniswap, we have all trade data which can be observed from the ethereum blockchain. We observe the pool for WETH/USDC, which is wrapped ethereum. The observation timeframe for the study the week of 1/1/2021 - 6/30/2021.

With this data we make two estimations. The first part of our study estimates a jump process as described above using our 1 minute data from Binance. The second estimation is a distribution for $\delta$ with the Uniswap data. \\

\subsection{Price Dynamics}
We use Maximum Likelihood Estimation (MLE) to calibrate the price dynamics of the pool from 15 second reference price data. Estimating the MJD model by MLE requires a parameter for the maximum number of jumps. We assume a maximum of three jumps. 

The log-likelihood estimation we use assumes that:

\[
  \begin{split}
    r_{t+\Delta t}
    & \defeq \log \frac{p_{t+\Delta t}}{p_t}
    \\
    & =  \mu_D \Delta t
    + \sigma_D \sqrt{\Delta t} Z_0
    + \sum_{i=1}^N \left( \mu_J + \sigma_J Z_i \right)
    \\ =
    &\mu_D \Delta t
    + N \mu_J
    + \sqrt{\sigma_D^2 \Delta t + N \sigma_J^2} Z.
  \end{split}
\]
Then,
\[
  \begin{split}
    \PR(r_{t+\Delta t} = r)
    &  \approx 
    \sum_{j=0}^{N_{\max}} \PR(N=k) \PR(r_{t+\Delta t} = r | N=k)
    \\
    &  \approx 
    \sum_{j=0}^{N_{\max}} \frac{(\lambda \Delta t)^k e^{-\lambda \Delta t}}{k!}
    \phi\left( r |  \mu_D \Delta t + N \mu_J, \sigma_D^2 \Delta t + N \sigma_J^2\right),
  \end{split}
\]
where $\phi(\cdot|\mu,\sigma^2)$ is the pdf of $N(\mu,\sigma^2)$.

The parameters we fit are below. We are using constant volatility over a long time frame (6 months). This is just a numerical calibration and in practice volatility is stochastic. We also note that other parameters are also non-stationary and in practice parameters would be updated more frequently. For robustness we set $\mu_d, \mu_j, \mu_\delta$ to 0, consistent with risk-neutral pricing. 

For clarity we interpret the variables as $\sigma_d$ is the standard deviation outside of jumps, $\lambda_j$ is the expected number of jumps per day and $\sigma_j$ is the expected movement of a jump. For NT1 the  $\sigma_\mu$ is the and the $\sigma_\delta$ is the standard deviation of the effective fair value around the actual value. For NT2 $\mu$ is the average size they trade at, $\lambda_{nt}$ is a parameter that controls for the shape of the exponential distribution of NT2 sizes and $rate_{nt}$ is the number of times per day that an NT2 arrives to trade.
 \\

\begin{table}[h]
    \begin{subtable}[h]{0.45\textwidth}
        \centering
        \begin{tabular}{c | c | l}
      Parameter & Estimated Value  \\
      \hline
      $\sigma_d$ & 0.080128 \\
      $\lambda_j$ & 1.070119 \\
      $\sigma_j$ & 0.013545  \\
       \end{tabular}
       \caption{MLE calibration of Merton Jump Diffusion (MJD) Process}
       \label{tab:week1}
    \end{subtable}
    \hfill
    \begin{subtable}[h]{0.45\textwidth}
        \centering
        \begin{tabular}{c | c | l}
        Parameter & Average Value \\
        \hline
         $reserve_0$ & 1.423880e8  \\
         $reserve_1$ & 7.837622e4 \\
         $\hat p_{initial}$ & reserve0 / reserve1 \\
        \end{tabular}
        \caption{Starting Value of the Uniswap Pool}
        \label{tab:week2}
     \end{subtable}
     \label{tab:temps}
\end{table} \

\begin{table}[h]
    \begin{subtable}[h]{0.45\textwidth}
        \centering
        \begin{tabular}{c | c | l}
      Parameter & Estimated Value  \\
      \hline
       $\lambda_\delta$ & 955.552632 \\
       $\sigma_\delta$ & 0.004034 \\
       \end{tabular}
       \caption{Parameters for the first type of noise trader (NT1)}
       \label{tab:week1}
    \end{subtable}
    \hfill
    \begin{subtable}[h]{0.45\textwidth}
        \centering
        \begin{tabular}{c | c | l}
        Parameter & Average Value \\
        \hline
         $\mu_{nt}$ & 14096 \\
         $\lambda_{nt}$ & 5.149200e-4 \\
         $rate_{nt}$ & 4,891 / day  \\
        \end{tabular}
        \caption{Parameters for the second type of noise trader (NT2)}
        \label{tab:week2}
     \end{subtable}
     \label{tab:temps}
\end{table}

\subsection{Pool Dynamics}
To calibrate the pool dynamics we do two things. First, we simply count number of pool transactions per unit time to determine the total arrival rate of actual trades. This doesn't correspond directly to the arrival rates in our model, since there is censoring; an agent that arrives with $\gamma p_t <= \tilde p_t <= \gamma^{-1} p_t$ does no observable trade. We address this problem by calibrating in simulation so that the simulated arrival rate of non-zero trades matches the observed total arrival rate.

The estimation for delta, the traders idiosyncratic value adjustment, is more involved. At the instant we observe a trade, we interpolate what the reference price $p_t$ would have been at that time. We also invert the Uniswap transition function equations above to back out the value of $\tilde p_t$ that's consistent with the observed reference trade. $\delta$ distribution: Given a pool trade $(u_x,u_y)$, we figure out unique $\tilde p_t$ consistent with that trade, and then $\delta = \tilde p_t / p_t - 1$.

The distributions of delta look like: \\

\begin{figure}[h!]
\centering
\includegraphics[width=1.0\textwidth]{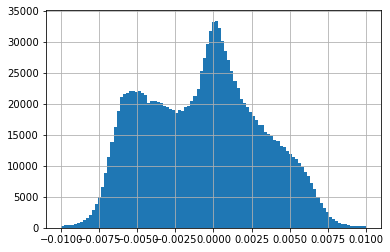}
\caption{Log ($\delta$) interpolated}
\end{figure}

During calibration we observe that the first type of noise trader (NT1), that trades according to their internal fair value, which is derived from a distribution around the true fair value does not fit the data very well. On the other hand we observe that the second type of noise trader (NT2) who picks a random direction and size to trade actually fits the data quite well. Although this result might seem surprising, crypto markets are have a high percentage of retail traders and it has been anecdotally observed that they often trade for reasons that are not based on a fundamental valuation, which is inherently difficult in crypto assets. 

Note that NT2 traders trade more often, but for smaller average size, than NT1. The core problem with the NT1 fit is that fees ($\gamma$) means that to trade much at all $\sigma_\delta$ needs to be quite big, but then the distribution of max(0, abs($\delta$)-30bps) has too fat tails, so the NT1 model suggests very large trade sizes. With the fit parameters, for NT1, we see an average trade size of \$78,415, but only a total of 1753 trades / day, and an average pool PnL of 9.85bps/day. This lines up poorly with the actual 7601 trades/day with an average size of \$14,096, and somewhat overstates the PnL. Moreover, exploring the parameter space, we were not able to find parameters that gave a realistic number of trades, average trade size, and PnL.

\subsection{Accounting Method}
A contribution of our approach is the method used for accounting. The standard approach to accounting for PnL on a delta hedged portfolio would be track PnL from each trade and also adjust the hedging portfolio which realizes its own PnL and then add the two PnL amounts together.  We observe that under risk neutral pricing that the expected PnL of the delta hedged portfolio is always zero. Specifically we assume risk neutral pricing of Ethereum itself, which allows us to account for all PnL within each trade. 

We take each trade done and compute the pool PnL $V_{t+} - V_{t-}$ of that trade. Summing these over all trades gives a very low noise estimate of the total expected PnL. (This approach is equivalent to continuously delta-hedging the pool's portfolio). This approach is more numerically stable and also allows us to obtain a low noise estimate of the greeks, which we discuss below. 

\subsection{Equilibrium Distribution}
We observe that there is scale-invariance with respect to the underlying price in this model. Both the pool-implied price $\hat p \defeq $ reserve0 / reserve1 and the noise trader idiosyncratic price $\tilde p$ can be normalized by the underlying fair price $p$. A benefit of this approach is that both tend to some equilibrium distribution. There is no closed form for our full model, but we can still simulate and sample, which allows faster simulation of larger numbers of runs.

However, for some simpler models we can analytically solve for this equilibrium of $\hat p / p$. We'll assume only arb traders. Although not exact, the fee very well approximates a $0.3\%$ fee, and likewise immediately after any trade we'll have $\log{\hat p / p} = \pm 0.003$. If arbs trade instantaneously, this is a confined brownian motion, and the equilibrium is uniformly distributed on $[-0.003, 0.003]$. If they instead arrive via a poisson arrival process, then the equilibrium is uniform within $\pm 0.003$ and decays exponentially outside. The decay rate being constant and proportional to the square root of the arrival rate. For a more rigorous derivation of this finding please see \citet{Milionis2023Fees} who we corresponded with in the process of developing this solution. 


\section{Numerical Results}
In this section we discuss the simulation methods in more detail and report the numerical results obtained from the previously described models. There are several steps to the simulation process. 

First, we bootstrap sample the initial pool conditions from the empirical equilibrium distributions. This allows us to converge a little faster because we omit the transient error from the start.  The next part of the simulation concerns the arrival times of both types of traders (arbs and noise traders). We then simulate the diffusion process between the arrival times of each type of trader. For the noise traders, we must also sample the action (direction traded and size for NT2) after they arrive to trade in the pool. In these simulations time discretization is not necessary.  


\begin{table}[!h]
\begin{center}
{\small
\begin{tabular}{c|c}
\hline
Parameter & Estimated Value  \\
\hline
Empirical $\mu$ & 8.25bps per day \\
Simulated $\mu$ & 8.12bps/day $\pm$ .23bps/day \\
\end{tabular}
}
\end{center}
\label{table2}
\caption{Parameters for the second type of noise trader M2}
\end{table}

\begin{figure}[h!]
\centering
\includegraphics[width=1.0\textwidth]{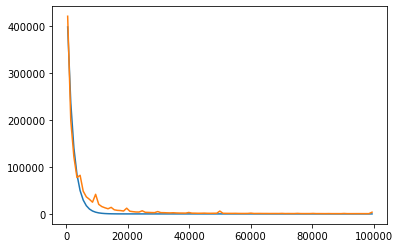}
\caption{Observed vs Model}
\end{figure}

\begin{figure}[h!]
\centering
\includegraphics[width=1.0\textwidth]{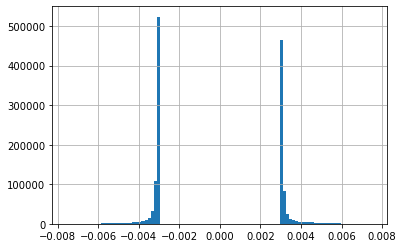}
\caption{log $\tilde p$ over pre-trade pool implied price}
\end{figure}

\begin{figure}[h!]
\centering
\includegraphics[width=1.0\textwidth]{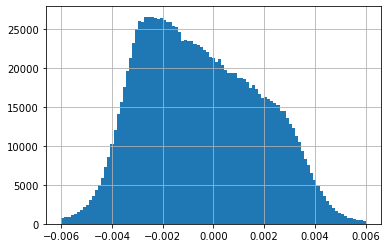}
\caption{log pool implied price over fair price}
\end{figure}

\section{Derivative Valuation}

A contribution of this work is examining the derivatives of various quantities. Understanding the sensitivities to various types of movements in the underlying instruments, fees and volatility is important for those market participants who want to hedge their position. The literature on derivatives is well developed in the traditional finance literature of options pricing. 

Many of the derivatives found in traditional finance don't have sensible interpretations in AMMs. For instance theta, which is the derivative with respect to expiry, doesn't have an AMM equivalent, because there is no expiry as is the case with an option\footnote {An analogy to theta for AMMs is the fees received on volume.}. The delta, which is the derivative with respect to price, does have an interpretation for our hedged portfolio and is functionally zero. 

Perhaps the most interesting of the standard derivatives to compute is gamma. We can think of participating in an AMM as having sold a limit option. This should have a negative gamma and in fact that is what we find as shown in Table 5. Moreover, this empirical finding corresponds with the theoretical results reported in \citet{Milionis2022LVR}. In some sense LVR can be thought of as an alternative way to measure gamma in the context of AMMs. In the LVR model of the world gamma should be exactly -1/4. We think this is an important insight and show the link between the two below:

Our formula is: \\
$\mathbb{E}(PnL) = \mathbb{E}(Volume)*$dt$ * fees + \delta* \mathbb{E}(Move) + .5*fees * \mathbb{E}(Squared Move)$ + higher order terms  \\

This formula is analogous to options pricing where the objective is to determine the value of an option as the underlying changes in price and time passes. In AMMs time isn't passing, so theta isn't relevant. Our formula can be used to determine the value when the underlying price changes using the same quadratic expansion as is done in the option pricing literature. More careful analysis would also take into account changes in expected future volatility, but that's outside the scope of this paper. 

The formula for LVR is: \\
$\mathbb{E}(PnL) = \mathbb{E}(Volume)*fee  - LVR$ \\

$LVR = (variance/8) * pool_value * dt$ \\

Using the numbers for our sample period this would yield: \\
$\mathbb{E}(PnL)$ = 15.73 - 8.27 = 7.46bps per day \\

In our experiments we find that our simulated gamma is 8.12 bps per day. We wouldn't expect them to align exactly as our gamma is naturally a more unstable and noisier estimate. Perhaps more importantly, this is only a second order approximation and higher order moments will matters, particularly since we are using non-normal distributions.    

We use the simulations to estimate more higher order sensitivities as one would do when pricing a path dependent options in the traditional finance literature. However CFMMs present their own set of interesting sensitivities, particularly those that correspond to the behavior of noise traders. These derivatives and their reported values are new contribution to the literature on AMMs. We report the findings of these derivatives in Table 5. 

One summary implication of these findings is that the fees should likely be set higher for AMM protocols. But this finding should be caveated by the idea that it assumes no change in trader behavior. it's likely that raising fees would reduce the amount of trading on the pool, but the impact of this would be difficult to measure without experimentation. 

The arb-trader-edge can be viewed as indicating that competition is good. More specifically, when there is more competition arb traders are forced to compete the edge down to the minimum which is good for the pool. the NT2 arrival rate is slightly larger than what a naive review of the PnL would indicate and likewise the size traded is slightly smaller. This indicates that all else equal, for the same increase in trading it would be preferable to have smaller more frequent trades, than less frequent, larger trades. \\

\begin{table}[!h]
\begin{center}
{\small
\begin{tabular}{c|c}
\hline
Parameter & Estimated Value  \\
\hline
$\delta$ & 0 \\
$\gamma$ & -.2498 \\
$\theta$ & avg trading rate * fees \\
\hline
d' wrt fees & 0.3451 (per 1bp increase in fee) \\
d' wrt arb-trader-edge & -.003576 (per 1bp change in trader edge) \\
d' wrt to NT2 arrival rate &  0.09775bps/day (per 1\% increase in NT2 rate) \\
d' wrt to NT2 size traded  &  0.07319bps/day  (per 1\% increase in NT2 size) \\
\end{tabular}
}
\end{center}
\label{table2}
\caption{Summary of Greeks}
\end{table}


\section{Conclusion}

In this paper we propose a simulation based model for valuing and hedging an LP share in Uniswap. We calibrated this model to observed data and model the behavior of agents based on centralized exchange prices and prices in the AMM pool. We find that on average LPs have an expected value 8.12bps per day on a hedged basis. We also determine the greeks of the pool and find that being an LP has negative gamma of -.25 as one would expect. Both of these results correspond with previous analytical findings, in particular LVR. 

A major caveat of this analysis is that it only examines a small window of time and Ethereum is a very volatile asset. It's entirely possible that this analysis would change significantly if we examined other time periods when the market structure and participants were different. Another limitation of this analysis is that we don't account for minting and burning, however we would not anticipate this having a material impact on our findings. We also note that the analysis was only done for one pool (V2, WETH-USDC) and that the ecosystem of traders and LPs has evolved since the time period we examine. In particular, a lot of liquidity has moved to Uniswap V3. Drawing implications for other pools and asset pairs is beyond the scope of this paper.   

We believe CFMMs such as Uniswap hold significant potential from both a research and practical perspective and recognize that these protocols are still very early in their development. This model should give those wanting to participate in CFMMs a better framework for thinking about proper hedging and the appropriate compensation required. 

Moreover, our model should be helpful for next generation protocol developers to think about alternative CFMM designs that may improve certain characteristics of the trading mechanism. Specifically, we view the ability to vary the LP fees as being a compelling market design feature. At an ecosystem level, our derivative estimations should point developers toward reducing gas fees and other frictions, to encourage more frequent trading. This allows users of CFMMs to trade at a lower cost and also benefits LP's who lose less to arbitrageurs. 

We see many lines of potential future research. Taking this model to data from other protocols such as Balancer and Sushi Swap and other large exchanges such as FTX and Coinbase would help confirm the results or perhaps uncover other phenomenon. Incorporating gas prices would be a logical next step. Gas prices are often variable and bound the amount of arbitrage activity. Building more complex models that explicitly include an arbitrageur who executes clean-up trades would also likely yield insights. 

AMMs represent a novel market design mechanism and their growing popularity suggest that they're here to stay. Our research helps explain some of the behavior between different types of agents that is observed and in AMMs such as Uniswap. We hope our empirical results and methods help other researchers trying to understand thee new market designs and refine better design for the next generation of AMMs. \\


\appendix

\section{Notes}

Merton jump-diffusion SDE:
\[
  \frac{dp_t}{p_t} =
  \left( \mu_D - \lambda_J k \right)\, dt
  + \sigma_D \, dB_t
  + (y_t - 1) \, dN_t,
\]
where $B_t$ is a standard Brownian motion, $N_t$ is the Poisson process with intensity
$\lambda_J$, and $\log y_t \sim N(\mu_J,\sigma_J^2)$. Here,
\[
k \defeq \E[y_t - 1] = \exp\left(\mu_J + \tfrac{1}{2} \sigma_J^2\right) - 1.
\]

Log returns:
\[
  \log \frac{p_{t+\Delta t}}{p_t}
  = \left( \mu_D - \tfrac{1}{2} \sigma_D^2\right) \Delta t
  + \sigma_D \sqrt{\Delta t} Z_0
  - \lambda k \Delta t
  + \sum_{i=1}^N \left( \mu_J + \sigma_J Z_i \right),
\]
where $N\sim \text{Poisson}(\lambda_J \Delta t)$, $Z_i \sim N(0,1)$, for $i \geq 0$.
Conditional on $N$, log returns are normally distributed, so that
\[
  \log \frac{p_{t+\Delta t}}{p_t}
  = \left( \mu_D - \tfrac{1}{2} \sigma_D^2- \lambda k\right) \Delta t
  + N \mu_J
  + \sqrt{\sigma_D^2 \Delta t + N \sigma_J^2} Z,
\]
where $Z \sim N(0,1)$.
Then,
\[
  p_{t+\Delta t} / p_t
  =
  \exp\left(
    \left( \mu_D - \tfrac{1}{2} \sigma_D^2- \lambda k\right) \Delta t
    + N \mu_J
    + \sqrt{\sigma_D^2 \Delta t + N \sigma_J^2} Z
  \right).
\]

Expected price:
\[
  \begin{split}
    \E\left[ p_{t+\Delta t} / p_t \right]
    & =
    \E\left[
      \E\left[ p_{t+\Delta t} / p_t | N \right]
    \right]
    \\
    &
    =
    \E\left[
      \exp\left(
        \left( \mu_D - \tfrac{1}{2} \sigma_D^2- \lambda k\right) \Delta t
        + N \mu_J
        + \tfrac{1}{2} \left( \sigma_D^2 \Delta t + N \sigma_J^2 \right)
      \right)
    \right]
    \\
    &
    =
    \E\left[
      \exp\left(
        \left( \mu_D - \lambda k\right) \Delta t
        + N \left(  \mu_J
          + \tfrac{1}{2} \sigma_J^2 \right)
      \right)
    \right]
    \\
    &
    = e^{
      \left( \mu_D - \lambda k\right) \Delta t
    }
    \E\left[
      \exp\left(
        N \left(  \mu_J
          + \tfrac{1}{2} \sigma_J^2 \right)
      \right)
    \right]
    \\
    &
    = e^{
      \left( \mu_D - \lambda k\right) \Delta t
    }
    \exp\left(
      \lambda \Delta t
      \left(  e^{\left( \mu_J
            + \tfrac{1}{2} \sigma_J^2 \right)}
        - 1
      \right)
    \right)
    \\
    &
    = e^{\mu_D \Delta t}.
  \end{split}
\]

Volatility:
\[
  \begin{split}
    \Var\left(  \log \frac{p_{t+\Delta t}}{p_t} \right)
    & =
    \E\left[
      \Var\left(  \left. \log \frac{p_{t+\Delta t}}{p_t} \right| N\right)
    \right]
    +
    \Var\left( \E\left[ \left.
          \log \frac{p_{t+\Delta t}}{p_t} \right| N \right] \right)
    \\
    & =
    \E\left[ \sigma_D^2 \Delta t + N \sigma_J^2 \right]
    +
    \Var\left(
      N \mu_J
    \right)
    \\
    & =
    \left( \sigma_D^2 + \lambda \left( \mu_J^2 + \sigma_J^2 \right) \right)\Delta t.
  \end{split}
\]

Log-likelihood (for estimation):
Assume that
\[
  \begin{split}
    r_{t+\Delta t}
    & \defeq \log \frac{p_{t+\Delta t}}{p_t}
    \\
    & =  \mu_D \Delta t
    + \sigma_D \sqrt{\Delta t} Z_0
    + \sum_{i=1}^N \left( \mu_J + \sigma_J Z_i \right)
    \\
    & = \mu_D \Delta t
    + N \mu_J
    + \sqrt{\sigma_D^2 \Delta t + N \sigma_J^2} Z.
  \end{split}
\]
Then,
\[
  \begin{split}
    \PR(r_{t+\Delta t} = r)
    & =
    \sum_{j=0}^{N_{\max}} \PR(N=k) \PR(r_{t+\Delta t} = r | N=k)
    \\
    & =
    \sum_{j=0}^{N_{\max}} \frac{(\lambda \Delta t)^k e^{-\lambda \Delta t}}{k!}
    \phi\left( r |  \mu_D \Delta t + N \mu_J, \sigma_D^2 \Delta t + N \sigma_J^2\right),
  \end{split}
\]
where $\phi(\cdot|\mu,\sigma^2)$ is the pdf of $N(\mu,\sigma^2)$.


{\small\singlespacing
  \bibliographystyle{plainnat}
  \bibliography{references}
}

\end{document}